\documentclass[hyper,letterpaper,12pt]{article}
\usepackage{a4wide}
\usepackage{amsmath}
\usepackage[
      colorlinks=true,
      linkcolor=blue,
      urlcolor=blue,
      filecolor=black,
      citecolor=blue,
            ]{hyperref}
\usepackage{color}
\usepackage{graphicx}
\usepackage{amsfonts}
\usepackage{amssymb}
\usepackage{epstopdf}

\begin{document}
\title{\bf \Large Thermodynamics of  Black Holes in Massive Gravity}

\author{\large
~Rong-Gen Cai$^1$\footnote{E-mail: cairg@itp.ac.cn}~,
~~Ya-Peng Hu$^2$\footnote{E-mail: huyp@nuaa.edu.cn}~,
~~Qi-Yuan Pan$^3$\footnote{E-mail: panqiyuan@126.com}~,
~~Yun-Long Zhang$^1$\footnote{E-mail: ylzhang@itp.ac.cn}
\\
\\
\small $^1$State Key Laboratory of Theoretical Physics,
\small Institute of Theoretical Physics, \\
\small Chinese Academy of Sciences,
\small Beijing 100190,  China\\
\small $^2$ College of Science, Nanjing University of Aeronautics \\
\small and Astronautics, Nanjing 211106, China \\
\small  $^3$ Institute of Physics and Department of Physics,\\
\small Hunan Normal University, Changsha, Hunan 410081, China}
%\date{\today}
\maketitle

\begin{abstract}
\normalsize
We present a class of charged black hole solutions in an ($n+2)$-dimensional massive gravity with a negative cosmological constant, and study  thermodynamics and phase structure of the black hole solutions both in grand canonical ensemble and canonical ensemble. The black hole horizon can have a positive, zero or negative constant curvature characterized by constant $k$. By using Hamiltonian approach, we obtain conserved charges of the solutions and find black hole entropy still obeys the area formula and the gravitational field equation at the black
hole horizon can be cast into the first law form of black hole thermodynamics. In grand canonical ensemble, we find that thermodynamics and phase structure depends on the combination $k -\mu^2/4 +c_2 m^2$ in the four dimensional case, where $\mu$ is the chemical potential and $c_2m^2$ is the coefficient of the second term in the potential associated with graviton mass. When it is positive, the Hawking-Page phase transition can happen, while as it is negative, the black hole is always thermodynamically stable with a positive capacity.  In canonical ensemble,  the combination turns out to be $k+c_2m^2$ in the four dimensional case. When it is positive, a first order phase transition can happen between small and  large black holes if the charge is less than its critical one. In higher dimensional  ($n+2 \ge 5$) case, even when the charge is absent, the small/large black hole phase transition can also appear, the coefficients for the third ($c_3m^2$) and/or the fourth ($c_4m^2$) terms in the potential associated with graviton mass in the massive gravity can play the same role as the charge does in the four dimensional case.

\end{abstract}

\tableofcontents

\section{ Introduction}
In 1983, Hawking and Page~\cite{Hawking:1982dh} found that there is a phase transition between Schwarzschild AdS black hole and thermal gas in anti-de Sitter (AdS) space. Nowadays the phase transition is named Hawking-Page phase transition in the literature. However, similar phase transition does not exist for black holes in asymptotically flat or de Sitter spacetimes,  The main reason for the existence of the Hawking-Page phase transition is as follows. In AdS space, there exists a minimal temperature, below which there is no black hole solution, but a stable thermal gas solution exists, while above the minimal temperature, there
are two black hole solutions with a same temperature, the black hole with larger horizon radius is thermodynamically  stable with a positive heat capacity, and the black hole with smaller horizon radius is thermodynamically unstable with a negative heat capacity, behaving like the Schwarzschild black hole in asymptotically flat space-time.   Thus beyond the minimal temperature, the thermal gas in AdS space will collapse to form the stable large black hole. This is just the Hawking-Page phase transition. Due to the AdS/CFT correspondence~\cite{Maldacena:1997re,Gubser:1998bc,Witten:1998qj}, which says that a quantum gravity in AdS space is dual to a conformal field theory (CFT) living on the boundary of the AdS space, the Hawking-Page phase transition received  another interpretation in the dual CFT side, the confinement/deconfinement phase transition of the dual gauge field theory~\cite{Witten:1998zw}.

Another remarkable difference of black hole  in AdS space from its counterpart in flat or de Sitter space is that the black hole in AdS space could have a Ricci flat or hyperbolic horizon, besides the sphere horizon.  Those black holes are usually called topological black holes in the literature~\cite{Lemos:1994xp,Lemos:1995cm,Huang:1995zb,Cai:1996eg,Mann:1996gj,Brill:1997mf,Vanzo:1997gw} .  It is quite interesting to note that the Hawking-Page phase transition does not happen
for the AdS black holes with Ricci flat or hyperbolic horizon~\cite{Birmingham:1998nr} .

The Einstein's general relativity is a relativistic theory of gravity where the graviton is massless. A natural question is whether one can build a self-consistent gravity theory if  graviton is massive. It turns out that it is
not a trivial matter. 	In \cite{deRham:2010ik,deRham:2010kj,Hinterbichler:2011tt}, a classical of nonlinear massive gravity theories has been proposed, in which the ghost field is absent~\cite{Hassan:2011hr,Hassan:2011tf}.  In this class of massive gravity, the energy-momentum is on longer conserved due to the breakdown of diffeomorphism invariance.  Recently, Vegh~\cite{Vegh:2013sk} found a nontrivial black hole solution with a Ricci flat horizon  in four dimensional massive gravity with a negative cosmological constant~\cite{Hassan:2011vm}. He then found that the mass of the graviton can play the same role as the lattice does in the holographic conductor model:  the conductivity
generally exhibits a Drude peak which approaches to a delta function in the massless gravity limit. Some holographic consequence of the effect of graviton mass in massive gravity has been
investigated in~\cite{Blake:2013bqa,Davison:2013jba,Davison:2013txa}. The propose of this paper is to generalize Vegh's black hole solution and to study corresponding thermodynamical properties and phase structure of the black hole solutions.

The organization of this paper is as follows.  In Sec.~2 we present the exact charged black hole solutions with any horizon topology in an $(n+2)$-dimensional massive gravity, while in Sec.~3
we obtain thermodynamical quantities associated with the solution, and show that they obey the first law of black hole thermodynamics  and the black hole entropy satisfies the area formula
as in general relativity (GR). In particular, although the massive gravity is not diffeomorphism invariant, the equation of motion of the gravitational field at a black hole horizon can be
caste into the first law form of black hole thermodynamics.   In Sec.~4 we study phase structure of a 4-dimensional black hole both in grand canonical ensemble  and canonical ensemble.
In Sec.~5 we discuss the case of a 5-dimensional neutral black hole and show that there exists a first order phase transition between small/large black holes, although in this case the electric charge
is absent.   We end this paper with conclusions in Sec.~6.\footnote{ While this work was preparing, the paper \cite{Adams:2014vza} appeared in the archive, the authors investigated the Hawking-Page phase transition in a four-dimensional neutral  black hole in a special class of massive gravity (with $c_2m^2$ term in this paper) and found that the Hawking-Page phase transition exists even
temperature goes down to zero.}

%_________________________________________________________________%

\section{The black hole solution}
\label{sect:solution}
Let us consider the following action for an $(n+2)$-dimensional  massive gravity~\cite{Vegh:2013sk}
\begin{equation}
\label{eq1}
S =\frac{1}{2\kappa^2}\int d^{n+2}x \sqrt{-g} [ R +\frac{n(n+1)}{l^2} -\frac{1}{4} F^2 +m^2 \sum^4_i c_i {\cal U}_i (g,f)],
\end{equation}
where   $f$ is a fixed symmetric tensor and usually is called the reference metric,
$c_i$ are constants,\footnote{For a self-consistent massive gravity theory, all those coefficients  might be required to be negative if $m^2>0$. However, in this paper we do not impose this limit, since in AdS space, the fluctuations of some fields with negative mass square could  still be stable if the mass square
obeys corresponding Breitenlohner-Freedman bounds. }  and ${\cal U}_i$ are symmetric polynomials of the eigenvalues of the $(n+2)\times (n+2)$ matrix ${\cal K}^{\mu}_{\ \nu} \equiv \sqrt {g^{\mu\alpha}f_{\alpha\nu}}$:
\begin{eqnarray}
\label{eq2}
&& {\cal U}_1= [{\cal K}], \nonumber \\
&& {\cal U}_2=  [{\cal K}]^2 -[{\cal K}^2], \nonumber \\
&& {\cal U}_3= [{\cal K}]^3 - 3[{\cal K}][{\cal K}^2]+ 2[{\cal K}^3], \nonumber \\
&& {\cal U}_4= [{\cal K}]^4- 6[{\cal K}^2][{\cal K}]^2 + 8[{\cal K}^3][{\cal K}]+3[{\cal K}^2]^2 -6[{\cal K}^4].
\end{eqnarray}
The square root in ${\cal K}$ means $(\sqrt{A})^{\mu}_{\ \nu}(\sqrt{A})^{\nu}_{\ \lambda}=A^{\mu}_{\ \lambda}$ and $[{\cal K}]=K^{\mu}_{\ \mu}$.  The equations  of motion turns out to be
\begin{eqnarray}
R_{\mu\nu}-\frac{1}{2}Rg_{\mu\nu}-\frac{n(n+1)}{2l^2} g_{\mu\nu}-\frac{1}{2}(F_{\mu\sigma}{F_{\nu }}^{\sigma}-\frac{1}{4}g_{\mu\nu}F^2)+m^2 \chi_{\mu\nu}&=&0,\nonumber \\
\nabla_{\mu} F^{\mu \nu}&=&0.~~
\end{eqnarray}
where
\begin{eqnarray}
&& \chi_{\mu\nu}=-\frac{c_1}{2}({\cal U}_1g_{\mu\nu}-{\cal K}_{\mu\nu})-\frac{c_2}{2}({\cal U}_2g_{\mu\nu}-2{\cal U}_1{\cal K}_{\mu\nu}+2{\cal K}^2_{\mu\nu})
-\frac{c_3}{2}({\cal U}_3g_{\mu\nu}-3{\cal U}_2{\cal K}_{\mu\nu}\nonumber \\
&&~~~~~~~~~ +6{\cal U}_1{\cal K}^2_{\mu\nu}-6{\cal K}^3_{\mu\nu})
-\frac{c_4}{2}({\cal U}_4g_{\mu\nu}-4{\cal U}_3{\cal K}_{\mu\nu}+12{\cal U}_2{\cal K}^2_{\mu\nu}-24{\cal U}_1{\cal K}^3_{\mu\nu}+24{\cal K}^4_{\mu\nu})
\end{eqnarray}

We are now looking for static black hole solution with metric
\begin{equation}
\label{metric}
ds^2 = - N^2(r) f(r) dt^2 + f^{-1}(r) dr^2 + r^2 h_{ij} dx^i dx^j , \ \  i, j =1,\ 2,\ 3, \cdots, n
\end{equation}
where $h_{ij}dx^idx^j$ is the line element for an Einstein space with constant curvature $n(n-1)k$.  Without loss of generality, one may take $k=1$, $0$, or $-1$,  corresponding to
a sphere, Ricci flat, or hyperbolic horizon for the black hole, respectively.   Following and generalizing the ansatz in \cite{Vegh:2013sk},  we take the following  reference metric as
\begin{equation}
\label{reference}
f_{\mu\nu} = {\rm diag}(0,0, c_0^2 h_{ij} )
\end{equation}
with $c_0$ being a positive constant.
%In the metric (\ref{metric}), the Maxwell field equations have the solution
%\begin{equation}
%\label{Maxwell}
%F^{tr}= \frac{q r^{n+2}}{h L^{n+2}},
%\end{equation}
%where $q$ is an integration constant.
With the  reference metric (\ref{reference}), we have
\begin{eqnarray}
\label{u}
&& {\cal U}_1= n c_0/r, \nonumber \\
&& {\cal U}_2=  n(n-1) c_0^2 /r^2, \nonumber \\
&& {\cal U}_3= n(n-1)(n-2) c_0^3/r^3, \nonumber \\
&& {\cal U}_4= n(n-1)(n-2)(n-3) c_0^4/r^4.
\end{eqnarray}
We see that in the 4-dimensional case with $n=2$, one has identically ${\cal U}_3={\cal U}_4=0$, while ${\cal U}_4=0 $  in the 5-dimensional case with $n=3$.

For the metric (\ref{metric}), the Hamiltonian action for the gravitational part turns out to be
\begin{equation}
\label{I_grav}
I_{\rm  Grav}   = -\frac{n V_n}{2\kappa^2}\int dt d r N U',
\end{equation}
where a prime denotes  the derivative with respect to $r$ and
$$
U= r^{n-1} (k-f) + \frac{r^{n+1}}{l^2} +m^2 \left ( \frac{c_0c_1}{n} r^n +c_0^2c_2 r^{n-1} + (n-1) c_0^3c_3 r^{n-2}+(n-1)(n-2)c_0^4 c_4 r^{n-3}\right)
   $$
 where $V_n$ is the volume of space spanned by coordinates $x_i$.  On the other hand, considering a static charged solution, it is easy to see that the Hamiltonian action for the Maxwell field part is of the form
 \begin{equation}
 \label{I_max}
 I_{\rm Max}= \frac{V_n}{2\kappa^2} \int dt dr \left [ \frac{ N}{2 r^n} p^2 + V p'\right ],
 \end{equation}
 where $p$ is the conjugate momentum of $A_r$ and $V=-A_t$.  Combining (\ref{I_grav}) and (\ref{I_max}), we have the total Hamiltonian action of the
 system
 \begin{equation}
 \label{I_total}
 I_{\rm total}= -\frac{nV_n}{2\kappa^2} \int dt dr \left [ N  (U'-\frac{1}{2n r^n}p^2)
 - \frac{1}{n}V  p'\right ].
 \end{equation}
 Varying the action, we have the equations of motion
 \begin{eqnarray}
 \label{EQ}
 && U' -\frac{1}{2n r^n}p^2=0, \nonumber \\
 && N'=0, \ \  p'=0,  \nonumber \\
 && V'=  \frac{N}{r^n} p,
 \end{eqnarray}
 Integrating the above equations, we have
 \begin{eqnarray}
 && N(r)=N_0, \ \  p = q, \nonumber \\
 && V = \mu - \frac{N_0}{(n-1) r^{n-1}}  q, \nonumber\\
 && U(r) = m_0 - \frac{1}{2n(n-1)r^{n-1}}  q^2,
 \end{eqnarray}
 where $N_0$, $q$, $\mu$ and $m_0$ are all constants. Without loss of generality one can set $N_0=1$ by rescaling the coordinate $t$. To have a vanishing static electric potential at a black hole  horizon $r_+$, we have the chemical potential at the infinity
 \begin{equation}
 \mu = \frac{1}{(n-1)r_+^{n-1}}q.
 \end{equation}
  As a result, we have the metric function $f(r)$ as
   \begin{eqnarray}
   \label{solution1}
   f(r) &=& k +\frac{r^2}{l^2} -\frac{m_0}{r^{n-1}}+\frac{q^2 }{2n(n-1)r^{2(n-1)} } + \frac{c_0c_1m^2}{n }r+c_0^2c_2m^2  \nonumber \\
       &&   +\frac{(n-1)c_0^3c_3m^2}{r}
 +\frac{(n-1)(n-2)c_0^4c_4m^2}{r^2} ,
   \end{eqnarray}
   expressed in terms of mass parameter $m_0$ and electric charge $q$,  or
 \begin{eqnarray}
   \label{solution}
      f(r) &=& k +\frac{r^2}{l^2} -\frac{m_0}{r^{n-1}}+\frac{(n-1) \mu^2 r_+^{2(n-1)}}{2nr^{2(n-1)} } +  \frac{c_0c_1m^2}{n}r +c_0^2c_2m^2 \nonumber \\
       &&   +\frac{(n-1)c_0^3c_3m^2}{r}
   +\frac{(n-1)(n-2)c_0^4c_4m^2}{r^2} ,
   \end{eqnarray}
   expressed in terms of the mass parameter $m_0$ and chemical potential.
 We see that in the 4-dimensional case, taking $c_3=c_4=0$ and $k=0$, one recovers the solution found in Ref.~\cite{Vegh:2013sk}.\footnote {Note that  here the coordinate $r$
 is different from the one in \cite{Vegh:2013sk}: $r_{\rm here}= l/r_{\rm there}$. }

\section{Thermodynamical quantities and the first law}

In this section we calculate conserved charges associated with the black hole solution found in the previous section. Note that the action density (\ref{I_total}) can be written as
\begin{equation}
I  = -\frac{n V_n (t_2-t_1)}{2\kappa^2}\int  d r \left [ N \left (U'-\frac{1}{2n r^{n}}p^2\right)
 - \frac{1}{n}V p'\right ]  +B,
\end{equation}
where $B$ is a surface term, which should be chosen so that the action has an extremum under variation of the fields with appropriate boundary conditions. One requires that the fields approach the classic solutions at infinity. Varying the action, one finds
the boundary term
\begin{equation}
\delta B= (t_2-t_1) (N_0\delta M - \mu \delta Q)+B_0 
\end{equation}
The boundary term $B$ is the conserved charge associated with the ``improper gauge transformation" produced by time evolution. The constant $B_0$ is determined by some physical consideration, for example, the mass vanishes when a black hole horizon goes to zero. Here $M$ and $N_0$ are a conjugate pair, and $\mu$ and $Q$ are another conjugate pair.  According to the Hamiltonian approach, we have the mass  $M$ and charge  $Q$ as
\begin{eqnarray}
\label{charge}
\label{mass}
 && M= \frac{n V_n}{2\kappa^2}m_0, \nonumber \\
&& Q= \frac{V_n}{2\kappa^2} q= \frac{(n-1) r_+^{n-1}V_n}{2\kappa^2 }\mu.
\end{eqnarray}
The black hole horizon is determined by $f(r)|_{r=r_+}=0$. Thus the mass  $M$ can be expressed in terms of the horizon radius $r_+$
\begin{eqnarray}
\label{Mass}
M &=& \frac{n V_nr_+^{n-1} }{2\kappa^2} \left ( k +\frac{r^2_+}{l^2}+\frac{q^2 }{2n(n-1)r_+^{2(n-1)} } + \frac{c_0c_1m^2 }{n }r_++c_0^2c_2m^2  \right.  \nonumber \\
       &&   +\frac{(n-1)c_0^3c_3m^2 }{r_+}
   \left. +\frac{(n-1)(n-2)c_0^4c_4m^2 }{r^2_+} \right)
    \end{eqnarray}
The Hawking temperature of the black hole can be easily obtained as
\begin{eqnarray}
\label{Temp}
T &=& \frac{1}{4\pi r_+} \left ( (n-1)k + (n+1)\frac{r_+^2}{l^2}  -\frac{q^2}{2nr_+^{2(n-1)}}  +c_1c_0 m^2  r_+ + (n-1)c_2c_0^2 m^2  \right.
\nonumber \\
 && +\left. \frac{(n-1)(n-2) c_3 c_0^3 m^2 }{r_+} +\frac{(n-1)(n-2)(n-3) c_4 c_0^4 m^2 }{r_+^2}\right ),
\end{eqnarray}
by requiring the Euclidean time ($\tau =i t$) has a period $\beta =4\pi/f'(r)|_{r=r_+} $ in order to remove the potential conical singularity at the black hole horizon and the
period just gives the inverse Hawking temperature of the black hole.
It is easy to  see that the black hole entropy obeys the area formula, which gives
\begin{equation}
\label{entropy}
S= \frac{2\pi V_n}{\kappa^2}r_+^n
\end{equation}
and then the following first law of black hole thermodynamics follows
\begin{equation}
d M = TdS +\mu dQ.
\end{equation}

Now let us notice an interesting properties of equations of motion of gravitational fields at a black hole horizon. Considering  the first equation in (\ref{EQ}),  it can be written as
\begin{eqnarray}
\label{em}
&& (n-1) r^{n-2} (k-f) - r^{n-1} f' + (n+1) \frac{r^n}{l^2} +m^2 ( c_0c_1 r^{n-1} +(n-1) c_0^2c_2r^{n-2} \nonumber \\
&&~~~~~ (n-1)(n-2) c_0^3c_3 r^{n-3} +(n-1)(n-2)(n-3) c_0^4 c_4 r^{n-4})= \frac{1}{2n r^n}q^2,
 \end{eqnarray}
 On the black hole horizon, one has $f(r)|_{r=r_+}=0$, and the black hole temperature
 \begin{equation}
 T= \frac{1}{4\pi} f'|_{r=r_+}.
 \end{equation}
 Considering  Eq.(\ref{em}) at the horizon,  multiplying $nV_n/(2\kappa^2) dr_+$ on both sides of (\ref{em}), we see that
 the equation can be rewritten as
 \begin{equation}
 dM-T d S  -\mu d Q=0,
 \end{equation}
 where $S$, $M$ and $Q$ are nothing, but the entropy in (\ref{entropy}), mass  and charge in (\ref{charge}). Thus we have shown
 that the Hamiltonian constraint, or say, the $t-t$ component of equations of gravitational field, on the horizon, can be cast into
 the first law form of black hole thermodynamics, like the case in GR, although the massive gravity manifestly breaks the diffeomorphism invariance, like the Horava-Lifshitz gravity. For the latter, we have also shown that the equation of motion of gravitational field at black hole horizon can be written as a form of the first law of black hole thermodynamics~\cite{CO}.

 \section{Phase structure of 4-dimensional black holes}

In this section we focus on the case in four dimensions. In that case, we have ${\cal U}_3={\cal U}_4=0$ ($c_3=c_4=0$) and the metric function $f(r)$ becomes
\begin{equation}
f(r) = k +\frac{r^2}{l^2} -\frac{m_0}{r}+\frac{q^2 }{4r^{2} } + \frac{c_1m^2}{2 }r+c_2m^2,
\end{equation}
where without loss of generality, we have set $c_0=1$.  Note that then vacuum solution with $m_0=q=0$  is
\begin{equation}
f_0(r)=  k +\frac{r^2}{l^2}  + \frac{c_1m^2}{2 }r+c_2m^2,
\end{equation}
which is not an AdS space unless $m^2=0$. Note that the black hole mass (\ref{mass}) is defined with respect to the vacuum solution.

\subsection{Grand canonical ensemble}
In a grand canonical ensemble with a fixed chemical potential $\mu$ associated with the charge $Q$, the Gibbs free energy is~\footnote{ Note that the Gibbs free energy (\ref{gibbs4})
is the same as  the Euclidean action difference of the black hole and corresponding vacuum solution multiplied by the black hole temperature.  }
\begin{eqnarray}
\label{gibbs4}
G_4 &=& M-TS-\mu Q, \nonumber \\
    &=& \frac{V_2r_+}{2\kappa^2}\left( k-\frac{r_+^2}{l^2}+c_2m^2 -\frac{1}{4}\mu^2\right).
\end{eqnarray}
The Hawking temperature can be written as
\begin{equation}
\label{temp4}
T_4=\frac{1}{4\pi r_+} \left( k+ 3\frac{r_+^2}{l^2} -\frac{1}{4}\mu^2 +c_1m^2r_+ + c_2 m^2 \right),
\end{equation}
and the heat capacity with a fixed chemical potential is given by
\begin{equation}
\label{capa4}
C_{\mu}= \left. T\left(\frac{dS}{dT}\right) \right |_{\mu}=\frac{(4\pi)^2 V_2 T r_+^3}{\kappa^2 (-k +3\frac{r_+^2}{l^2} +\frac{\mu ^2}{4} -c_2m^2)}.
\end{equation}
From these thermodynamical quantities, we see that in this grand canonical ensemble, the term $\mu^2- 4c_2m^2$ behaves as an effective chemical potential $\tilde {\mu}^2=\mu^2 -4c_2m^2$ for a charged black hole in GR. We now discuss the cases of $k=-1$, $0$ and $1$, respectively.

\subsubsection{The case of $k=-1$}

In this case, if $c_2 \le 0$, we see that $\tilde{\mu}^2$ is always positive, and the Gibbs free energy is always negative, \footnote{In this paper we always assume $m^2>0$.} while the Gibbs free energy  changes its sign at
\begin{equation}
r_{g c} = l\sqrt{-\tilde{\mu}^2/4 -1},
\end{equation}
if $c_2>0$ so that $-\tilde{\mu}^2 >4$. Namely, when $r_+ <r_{gc}$, one has $G_4>0$, while $G_4 <0$ as $r_+ >r_{g c}$. In other words, there does not exist the Hawking-Page
phase transition if $c_2 <0$, while it exists if $4c_2m^2-\mu^2 >4$. The Hawking-Page phase transition temperature is given by
\begin{equation}
T_{HP}=\frac{1}{4\pi r_{gc}} \left( 2\frac{r_{gc}^2}{l^2}+c_1m^2r_{gc}\right).
\end{equation}
To have a positive $T_{HP}$, we see that $c_1m^2$ has to satisfy: $c_1m^2 >-2 r_{gc}/l^2$.

The locally thermodynamical  stability of the black hole  is determined by the heat capacity (\ref{capa4}). We see that $C_{\mu}$ is always positive if $c_2 \le 0$; however, if $c_2 >0$ so that $-\tilde{\mu}^2>4$,  one has $C_{\mu}<0 $ when $r_+ <r_{\mu c}$, $C_{\mu}>0$ when $r_+>r_{\mu c}$, and it diverges
at
\begin{equation}
r_{\mu c}= l\sqrt{ (-\tilde{\mu}^2/4 -1)/3}=r_{gc}/\sqrt{3}.
\end{equation}
The divergence point corresponds to the minimal temperature (see the left plot of Fig.~\ref{T(k=-1)}).  We see that  in this case, the temperature behavior of the black hole with a hyperbolic
horizon is qualitatively the sam as the one of Schwarzschild AdS black hole in GR: there exists a minimal temperature, the smaller black holes with $r_+<r_{\mu c}$ are unstable with a
negative heat capacity, while the larger black holes with $r_+> r_{\mu c}$ are stable with a positive heat capacity.

In addition, we see from the temperature ({\ref{temp4}) that there exists a minimal horizon radius $r_m$
\begin{equation}
\label{mini-1}
r_m= \frac{c_1m^2l^2}{6} \left[ -1 \pm \sqrt{1 +\frac{12}{l^2 c_1^2m^4}\left(1+\frac{1}{4}\tilde{\mu}^2 \right) }\right ],
\end{equation}
where the Hawking temperature vanishes, it corresponds to an extremal black hole (see the right plot of Fig.~\ref{T(k=-1)}). Note that as $c_1 \ge 0$, one requires $1+\tilde{\mu}^2/4 \ge 0$, in order to have a real root $r_m\ge 0$ (in this case, one takes ``+" in (\ref{mini-1})). On the other hand, if $c_1 <0$, the condition has to be satisfied: $1+\tilde{\mu}^2/4 + l^2 c_1^2 m^4/12 \ge 0$, in this case, one takes ``$-$" in (\ref{mini-1}). Furthermore, if $c_1=0$,
the minimal horizon radius is $r_m= l\sqrt{(\tilde{\mu}^2/4+1)/3}$.

\begin{figure}[h]
\centering
\includegraphics[scale=0.7]{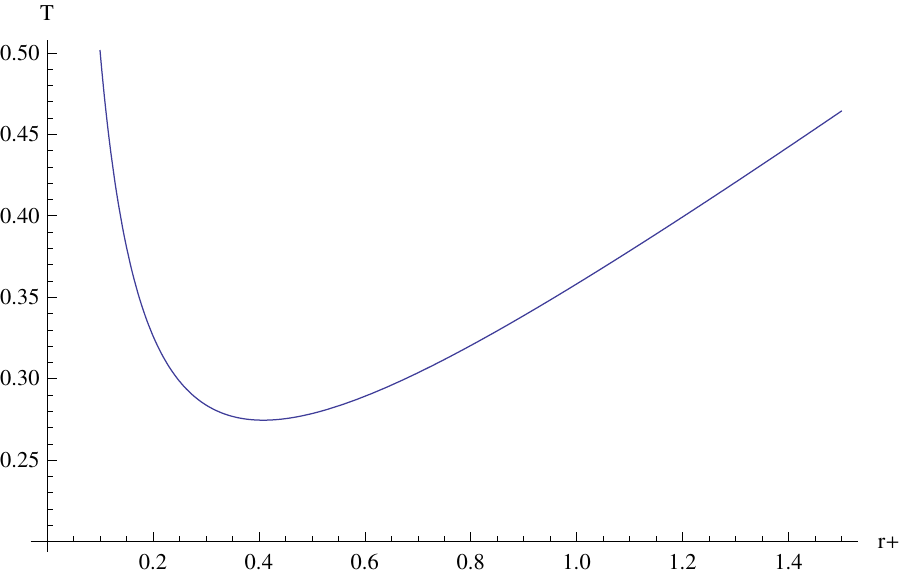}
\includegraphics[scale=0.7]{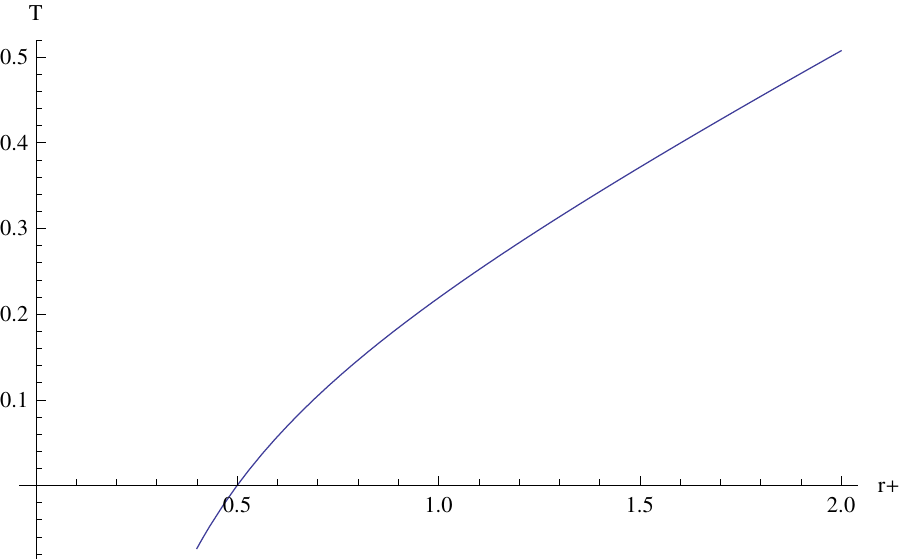}
\caption{The temperature $T$ of the black hole with $k=-1$ versus horizon radius $r_+$, here we take $l=c_1m^2=1$.  Left: $\tilde{\mu}^2=-6$. Right: $\tilde{\mu}^2=1$. }
\label{T(k=-1)}
\end{figure}

%\begin{figure}[h]
%\centering
%\includegraphics[scale=0.7]{T2(k=-1).pdf}\caption{The temperature of black hole with $k=-1$ versus horizon radius in the case of $k=-1$, here we take $l=1=c_1=1$.  $\tilde{\mu}=\sqrt{6}$. }
%\label{T2(k=-1)}
%\end{figure}

As a summary we find that  if both $c_1<0$ and $c_2<0$, the situation is qualitatively same as the case in GR: the black hole is not only globally thermodynamical stable with $G_4<0$, but also locally thermodynamical  stable with $C_{\mu}>0$; the minimal horizon radius (\ref{mini-1}) exists. However if $c_2>0$ so that $c_2m^2 >\mu^2/4$,  the thermodynamical behavior of the
black hole is qualitatively the same as the Schwarzschild AdS black hole: there does exist a minimal temperature, and Hawking-Page phase transition happens although the black hole discussed here has a hyperbolical horizon.

\subsubsection{The case of $k=0$}

In this case, the Gibbs free energy is always negative if $c_2 \le 0$. But when $c_2 >0$ so that $-\tilde{\mu}^2 >0$, the Gibbs free energy is positive for small horizon radius $r_+ < r_{gc}$, negative for $r_+>r_{gc}$ and changes its sign at
\begin{equation}
r_{gc}= l\sqrt{-\tilde{\mu}^2/4}.
\end{equation}
Namely if $c_2 >0$, the Hawking-Page phase transition can happen at $r_+=r_{gc}$.

The extremal black hole exists with a horizon radius
\begin{equation}
\label{mini0}
 r_m= \frac{c_1m^2l^2}{6} \left( -1 \pm \sqrt{1 +\frac{3}{l^2 c_1^2m^4}\tilde{\mu}^2 }\right).
\end{equation}
If $c_1 \ge 0$, we take ``+" in the above equation and require $\tilde{\mu}^2 >0$; if $c_1<0$, one takes ``$-$" and requires $ 1+3 \tilde{\mu}^2/(l^2 c_1^2 m^4)>0$;
and if $c_1=0$, the minimal horizon radius is $r_m=l\sqrt{\tilde{\mu}^2/12}$. See the left plot of Fig.~\ref{T(k=0)}.

\begin{figure}[h]
\centering
\includegraphics[scale=0.7]{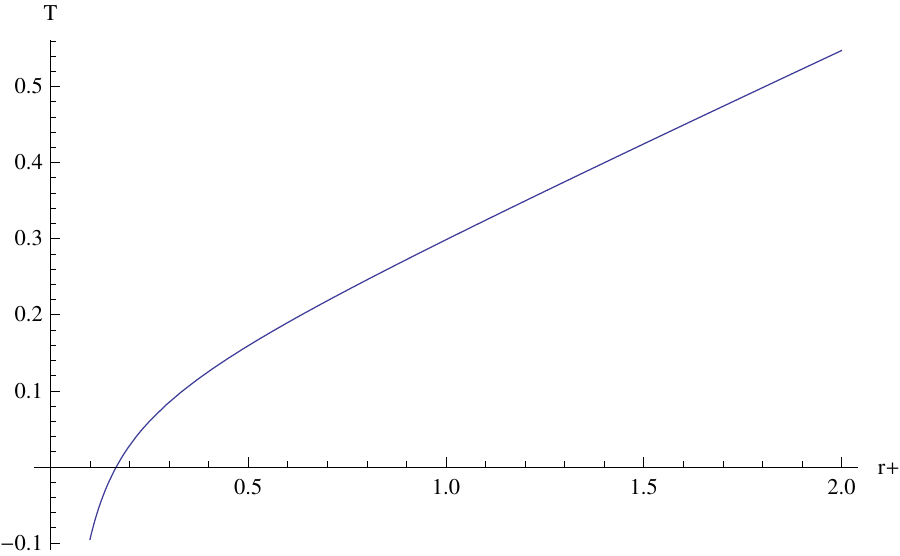}
\includegraphics[scale=0.7]{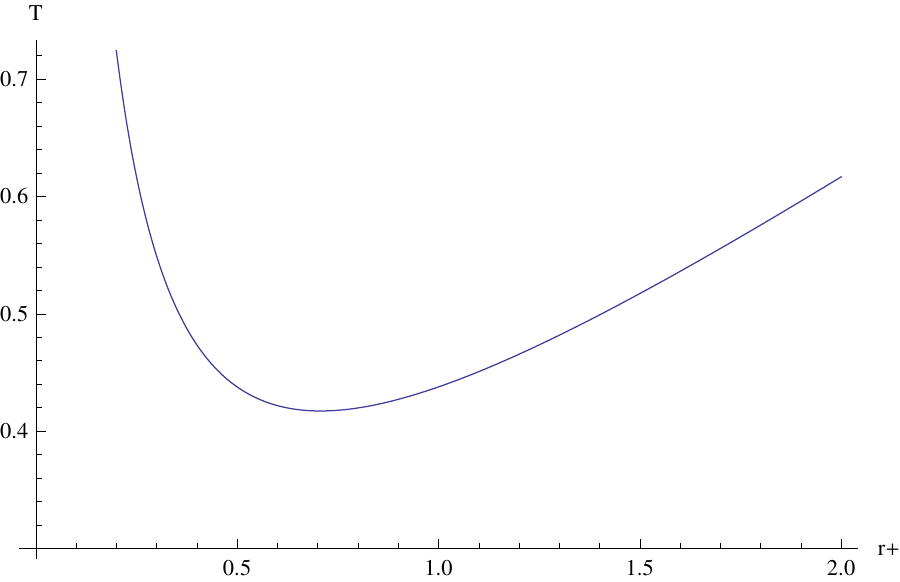}
\caption{The temperature $T$ of the black hole with $k=0$ versus horizon radius $r_+$, here we take $l=c_1m^2=1$.  Left: $\tilde{\mu}^2=1$. Right: $\tilde{\mu}^2=-6$. }
\label{T(k=0)}
\end{figure}

If $\mu <2 m\sqrt{c_2}$, there exists a minimal temperature at
\begin{equation}
r_{\mu c}= l\sqrt{ (c_2 m^2-\mu^2/4)/3}=r_{gc}/\sqrt{3}.
\end{equation}
When $r_+<r_{\mu c}$, the black hole has a negative heat capacity, and the heat capacity is positive as $r_+>r_{\mu c}$, and it diverges at $r_+=r_{\mu c}$. See the right plot of Fig.~\ref{T(k=0)}.

As a summary, if $c_1<0$ and $c_2<0$, the situation is qualitatively same as the case in GR: the black hole is not only global stable with $G_4<0$, but also local stable with $C_{\mu}>0$; the extremal black hole with the minimal horizon radius (\ref{mini0}) exists. However, if $c_2>0$ and so that $-\tilde{\mu}^2 >0$, once again,  the situation qualitatively behaves like the Schwarzschild AdS black hole, although now the black hole discussed here  has a Ricci flat horizon.

\subsubsection{The case of $k=1$}

In this case, we can see from (\ref{gibbs4}) that the Gibbs free energy changes its sign at
\begin{equation}
\label{rgc1}
r_{g c} = l\sqrt{1-\tilde{\mu}^2/4},
\end{equation}
which means that the Hawking-Page phase transition happens at $r_+=r_{gc}$~\cite{Adams:2014vza}, if $1-\tilde{\mu}^2/4 >0$. The Gibbs free energy is
positive for small black holes with $r_+ <r_{gc}$ and negative for large black holes with $r_+ >r_{gc}$. Of course, if $1-\tilde{\mu}^2/4 <0$, the Hawking-Page
phase transition does not appear.

From (\ref{capa4}) we see that if $1-\tilde{\mu}^2/4 >0$, there exists a minimal temperature at
\begin{equation}
r_{\mu c}= l\sqrt{ (1-\tilde{\mu}^2/4)/3}=r_{gc}/\sqrt{3}.
\end{equation}
When $r_+<r_{\mu c}$, the heat capacity is negative, while it is positive as $r_+>r_{\mu c}$. The heat capacity diverges at $r_+=r_{\mu c}$.  See the left plot of Fig.~\ref{T(k=1)}.

\begin{figure}[h]
\centering
\includegraphics[scale=0.7]{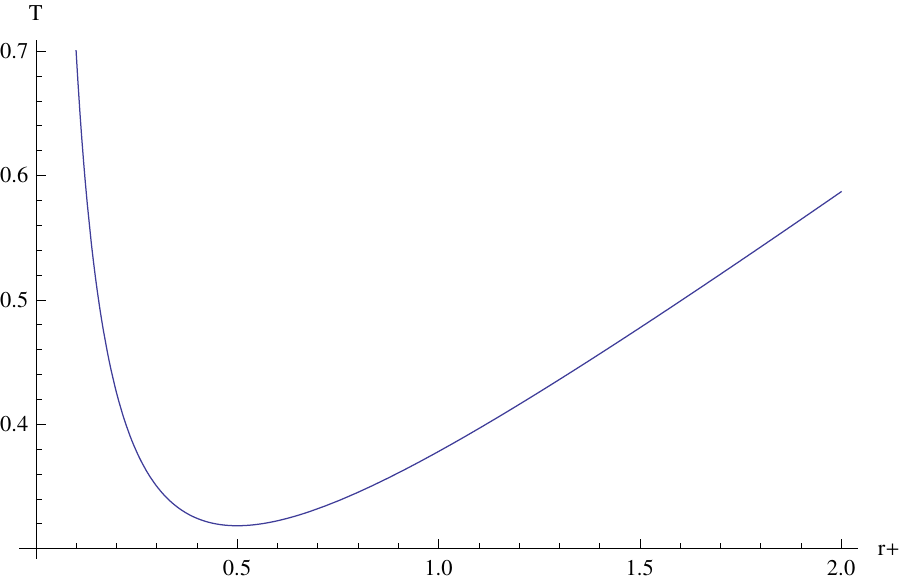}
\includegraphics[scale=0.7]{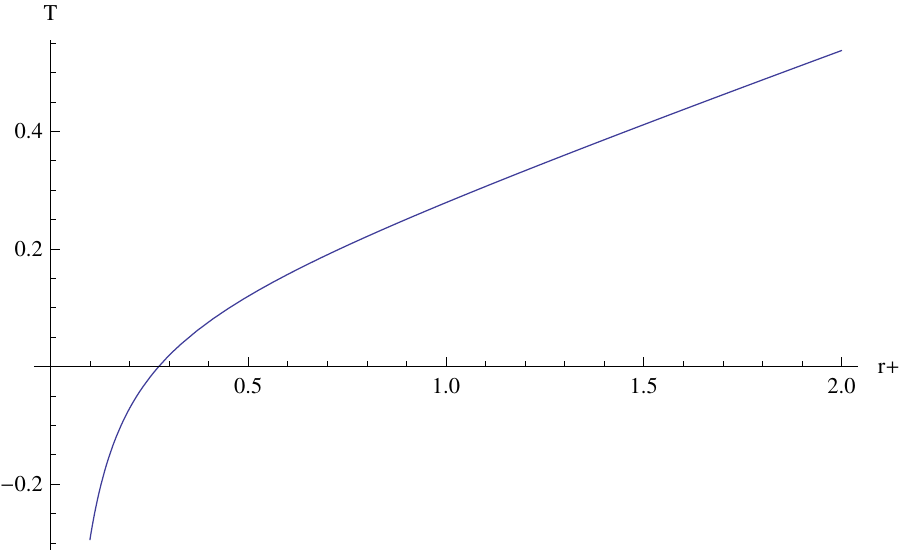}
\caption{The temperature $T$ of the black hole with $k=1$ versus horizon radius $r_+$, here we take $l=c_1m^2=1$.  Left: $\tilde{\mu}^2=1$. Right: $\tilde{\mu}^2=6$. }
\label{T(k=1)}
\end{figure}

On the other hand, if $1-\tilde{\mu}^2/4 <0$,  the extremal back hole solution with a vanishing Hawking temperature exists and its horizon radius is given by
\begin{equation}
\label{mini1}
r_m= \frac{l^2c_1m^2}{6} \left [ -1 \pm \sqrt{1 -\frac{12}{l^2 c_1^2m^4}\left(1-\frac{1}{4}\tilde{\mu}^2\right) }\right ].
\end{equation}
We see that if $c_1>0$, one takes the ``+" sign in (\ref{mini1});
if $c_1<0$, we take the ``$-$" sign; and if $c_1=0$,  we have
$r_m= l\sqrt{\tilde{\mu^2}/4-1}$.   See the right plot of Fig.~\ref{T(k=1)}.

Thus wee see when both $c_1<0$ and $c_2<0$, once again, the situation is qualitatively same as the case of Schwarzschild AdS black hole in GR: the black hole is locally unstable for mall black holes with $r_+<r_{\mu c}$, while it is stable for large black holes with $r_+>r_{\mu c}$; and the Hawking-Page phase transition happens at $r_+=r_{gc}$. The Hawking-Page phase transition
does not exists if the effective potential $\tilde{\mu}^2/4 >1$.

To summarize the three cases, we find that thermodynamics and phase structure of the black holes crucially depend on the combination $k-\mu^2/4+c_2m^2$. When it is positive, there is a minimal temperature, the black hole with smaller horizon are thermodynamically unstable, while it is stable for larger horizon, and the Hawking-Page phase transition can happen for any topological horizon ($k=1, 0,-1$). This is quite different from the case in GR, where the Hawking-Page phase transition can appear only in the case of $k=1$. When the combination is
negative, the black hole is always thermodynamically stable and no phase transition can happen.

%if both $c_1$ and $c_2$ are negative, the thermodynamical properties of the charged black holes in massive gravity is qualitatively same as %charged black holes in GR in a grand canonical ensemble. However, if $c_2$ is positive and so that the effective chemical potential satisfies %$k-\tilde{\mu}^2/4 >0$,  we find  that the Hawking-Page phase transition always can happen for any topological horizon ($k=1, 0,-1$) for %black holes in massive gravity. This is quite different from the case in GR, where the Hawking-Page phase transition can appear only in the %case of $k=1$. In fact, as we can see from those thermodynamical quantities (\ref{gibbs4}), (\ref{temp4}) and (\ref{capa4}), the term %$c_2m^2$ can also be
%combined into the horizon topology term $k$, so that the black hole solution looks like an effective horizon topology $\tilde k= k+c_2m^2$.  %In the following section, we will consider the black hole thermodynamics from this view of point in a canonical ensemble.

\subsection{Canonical ensemble}
In a  canonical ensemble with a fixed  charge $Q$, the Helmholtz free energy is~\footnote{If we define an extremal black hole with charge $q$ as the reference background~\cite{Chamblin:1999tk}, the Helmholtz free energy then should be
changed to $\tilde{F_4} = F_4 -M_{ext}$, where $M_{ext}$ is the mass of the extremal black hole with vanishing Hawking temperature. The conclusions are not qualitatively changed if one takes $F_4$ instead of $\tilde{F_4}$, therefore we here consider $F_4$ for simplicity.}
\begin{eqnarray}
\label{F4}
F_4 &=& M-TS, \nonumber \\
    &=& \frac{V_2r_+}{2\kappa^2}\left( k-\frac{r_+^2}{l^2} +c_2m^2 +\frac{3q^2}{4r_+^2}\right),
    \end{eqnarray}
 associated heat capacity is given by
\begin{equation}
\label{heat4}
C_Q=\left. T\left(\frac{dS}{dT}\right) \right |_Q=\frac{(4\pi)^2 V_2 T r_+^3}{\kappa^2 (-k +3\frac{r_+^2}{l^2} +\frac{3q^2}{4r_+^2} -c_2m^2)},
\end{equation}
and the black hole temperature can be expressed in terms of the horizon radius and charge as
\begin{equation}
T_4=\frac{1}{4\pi r_+} \left( k+ 3\frac{r_+^2}{l^2} -\frac{1}{4}\frac{q^2}{r_+^2} +c_1m^2r_+ + c_2 m^2 \right).
\end{equation}
We see from these thermodynamical quantities that we can combine $k$ and $c_2m^2$ as an effective horizon curvature $\tilde{k}=k+c_2m^2$.

From (\ref{F4}) we see that the Helmholtz free energy is positive for small black holes with $r_+ <r_{fc}$, and negative for large black holes with $r_+ >r_{fc}$ and changes its sign at $r_+=r_{fc}$,  where
\begin{equation}
r_{fc}^2 = \frac{l^2}{2} \left( \tilde {k} +\sqrt{\tilde{k}^2 +\frac{3q^2}{l^2}}\right).
\end{equation}
On the other hand, we see from the heat capacity (\ref{heat4}) that if $\tilde {k} \le 0$, the heat capacity is always positive.  However, if $\tilde{k}>0$, the heat capacity is positive for small black holes with $r_+ < r_{qc-}$ and large black holes with $r_+> r_{qc+}$, negative
for the intermediate black holes with $r_{qc-} <r_+ < r_{qc+}$, and diverges at
\begin{equation}
r_{qc\pm}^2= \frac{l^2}{6}\left( \tilde{k} \pm \sqrt{\tilde{k}^2 -\frac{9q^2}{l^2}} \right),
\end{equation}
if the charge is less than the critical one $q_{crit}^2= l^2 \tilde{k}^2/9$. Of course, if the charge is larger than the critical one, the
black hole is always stable with positive heat capacity. These properties can be seen from the behavior of the Hawking temperature (see Figs. \ref{T(q=1,k=0,-2)} and \ref{T(k=2)}).
Note that those properties are independent of the sign of $c_1$ as in the grand canonical ensemble.

\begin{figure}[h]
\centering
\includegraphics[scale=0.7]{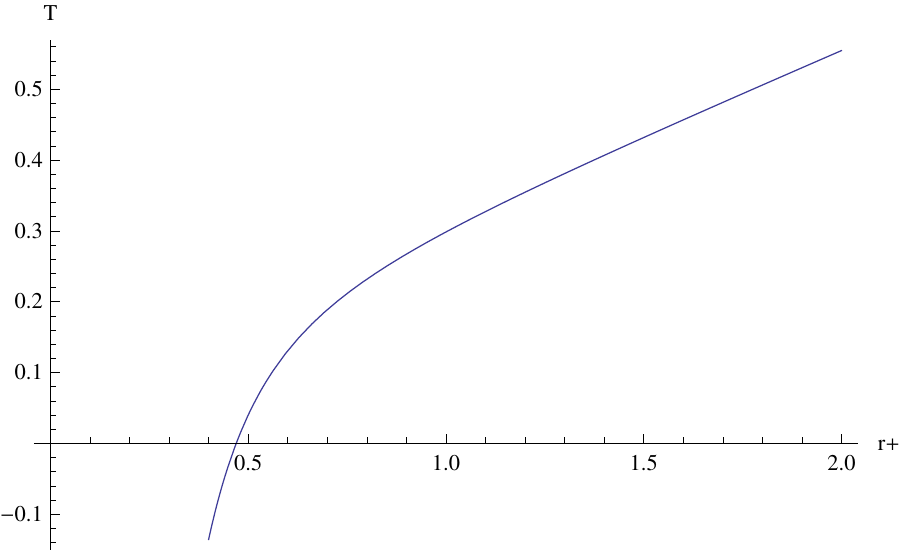}
\includegraphics[scale=0.7]{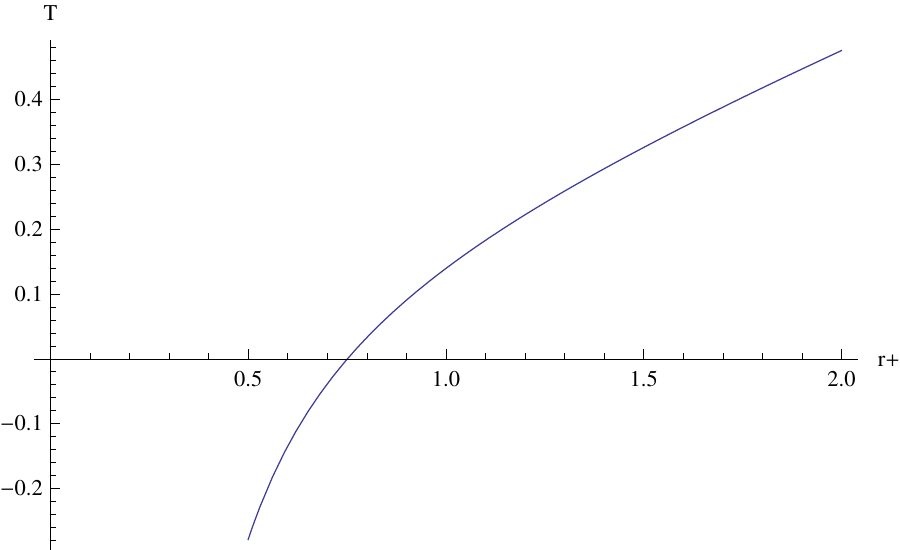}
\caption{The temperature $T$ of the black hole with $q =1$ versus horizon radius $r_+$, here we take $l=c_1m^2=1$.  Left: $\tilde{k}=0$. Right: $\tilde{k}=-2$. }
\label{T(q=1,k=0,-2)}
\end{figure}

\begin{figure}[h]
\centering
\includegraphics[scale=0.7]{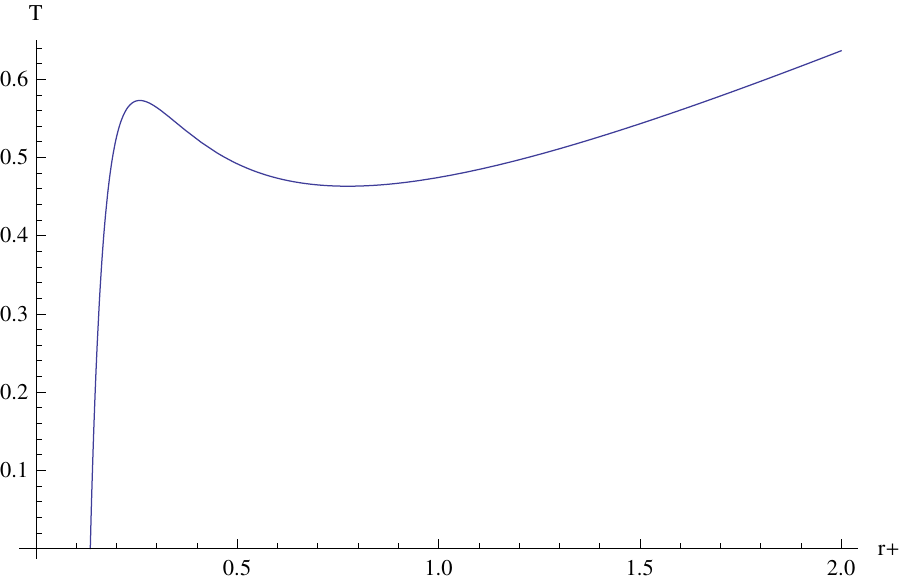}
\includegraphics[scale=0.7]{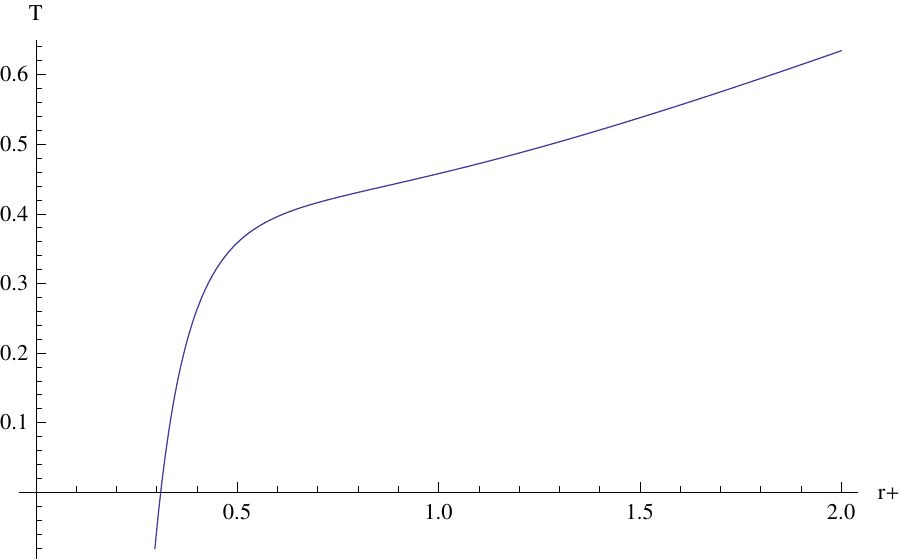}
\caption{The temperature $T$ of the black hole with $\tilde{k} =2 $ versus horizon radius $r_+$, here we take $l=c_1m^2=1$.  Left: $q=0.4$ ($< q_{crit}$). Right: $q=1$ ($>q_{crit}$). }
\label{T(k=2)}
\end{figure}

Note that the horizon radius for an extremal black hole with vanishing Hawking temperature is determined by the following equation
\begin{equation}
\tilde{k}+ 3\frac{r_m^2}{l^2} -\frac{1}{4}\frac{q^2}{r_m^2} +c_1m^2r_m =0,
\end{equation}
which has a simple root
\begin{equation}
r^2_{m}= \frac{l^2}{6}\left(-\tilde{k}+\sqrt{\tilde{k}^2 +\frac{3q^2}{l^2}}\right),
\end{equation}
if $c_1=0$.  Let us remind here that the minimal horizon radius always exists for any $\tilde{k}$.

In Fig.~\ref{F(k=2)}  we plot the Helmholtz free energy of the black holes with $\tilde{k}=2$  with respect to temperature in the cases
of $q<q_{crit}$ and $q>q_{crit}$, respectively.  In the left plot,~\footnote{Note that if one plots the free energy $\tilde{F}_4=F_4-M_{ext}$,  the free energy $\tilde{F}_4$ has a same shape as
$F_4$, the only difference is that the whole free energy curve $\tilde{F}_4$ will move to below the horizontal axes. Namely $\tilde{F}_4$ is always negative.} we see that a typical first order phase transition signature, a swallow tail, appears when
$q<q_{crit}$, while it disappears when $q>q_{crti}$.  Thus we see that as the case in GR, if $\tilde{k}>0$, there exists a first order phase transition between small and large black holes
if the charge is less than the critical one.  The phase transition disappears when the charge is larger than the critical one. The critical
point appears when the charge arrives at the critical one.  This phase transition behaves very like the one in the van der Waals system as in the case of
charged black holes in GR~\cite{Chamblin:1999tk}.   Note that such a phase transition does not exist for the case of $\tilde{k}\le 0$.

\begin{figure}[h]
\centering
\includegraphics[scale=0.5]{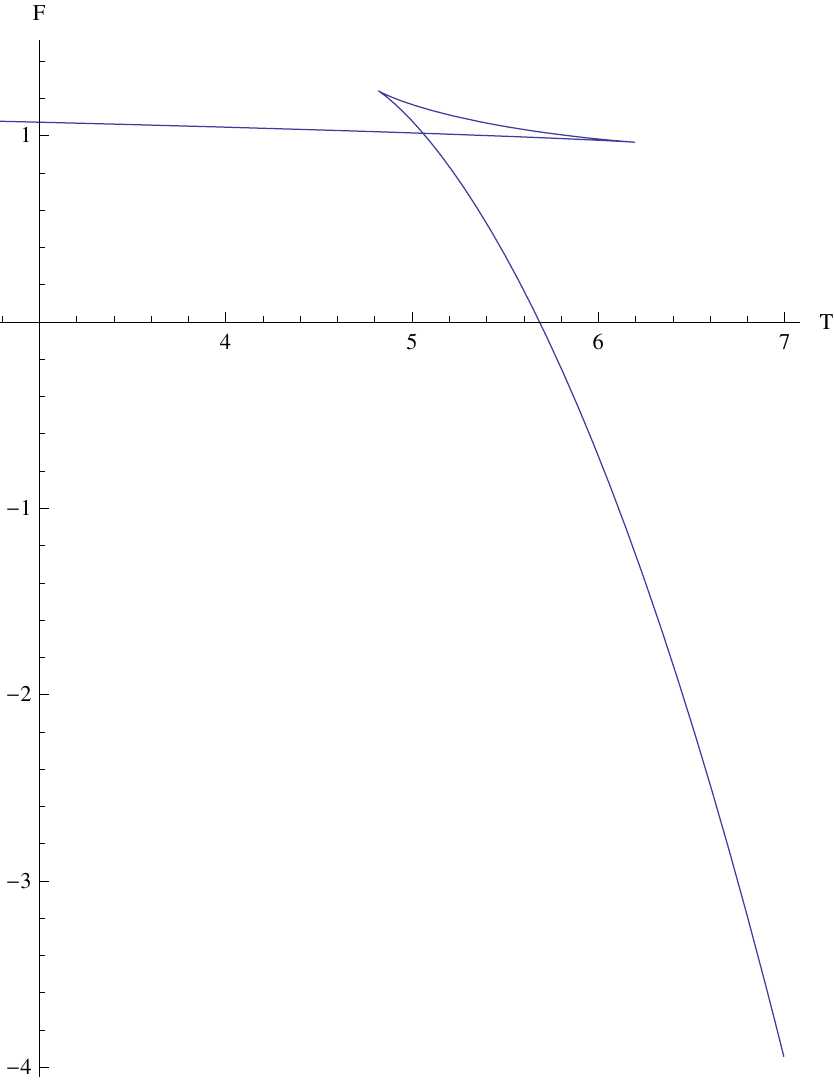}
\hspace{2.cm}
\includegraphics[scale=0.5]{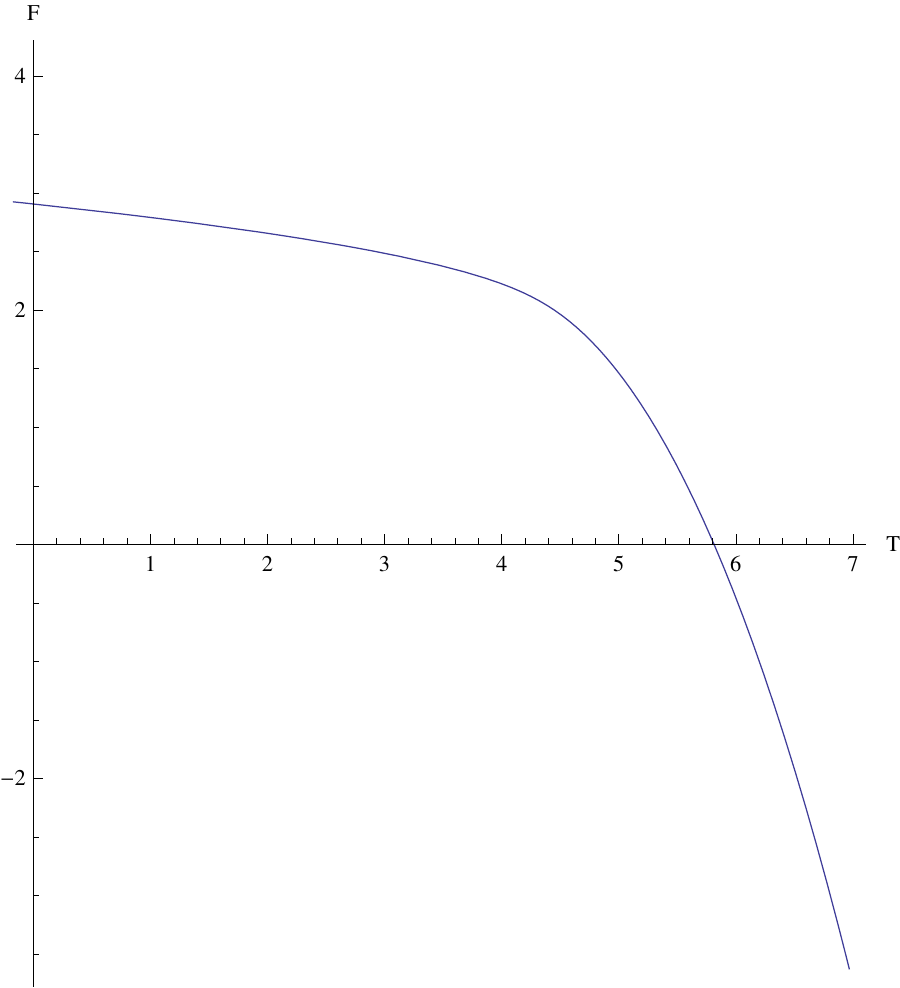}
\caption{The Helmholtz free energy of black hole  with $\tilde{k} =2 $ versus temperature $T$, here we take $l=1$, $c_1m^2=0$.  Left: $q=0.4$ ($< q_{crit}$). Right: $q=1$ ($>q_{crit}$). }
\label{F(k=2)}
\end{figure}

As a summary, we see that in canonical ensemble, the thermodynamical behavior of the charged black holes in massive gravity is qualitatively  same as that of Reissner-Nordstr\"om AdS
black holes with an effective horizon curvature $\tilde {k}= k +c_2m^2 $. If $\tilde{k}>0$, there exists a small/large black hole phase transition if the charge is less than the critical one. This
phase transition disappears when the charge is larger than the critical one.

\section{The phase transition in higher dimensional neutral black holes}

In a higher dimensional case with $n>2$,  one has $c_3 \ne 0$ if $n\ge 3$, and $c_3 \ne 0$ , $c_4\ne 0$ if $n\ge 4$, in general.  Note that the charge plays a curial role in the existence
of small/large black hole phase transition in the four dimensional case.  In this section, we show that  $c_3m^2$  and $c_4m^2$ terms can play the same role as the charge in higher dimensional cases.  Take an example, here  we consider the case of 5-dimensional neutral black holes. In this case,  to clearly see the effect of the term $c_3m^2$, we set $c_1=c_2=q=0$ foe simplicity in this section.  Thus the Hawking temperature is given by
\begin{equation}
T_5=\frac{1}{4\pi r_+} \left (2 k +4 \frac{r_+^2}{l^2}+ \frac{2c_3m^2}{r_+}\right).
\end{equation}
the mass of the black hole
\begin{equation}
M_5= \frac{3V_3 r_+^2}{2\kappa^2} \left( k+\frac{r_+^2}{l^2} +\frac{2c_3m^2}{r_+}\right),
\end{equation}
and the Helmholtz free energy
\begin{equation}
F_5= \frac{V_3r_+^2}{2\kappa^2} \left(k-\frac{r_+^2}{l^2} +\frac{4c_3m^2}{r_+}\right).
\end{equation}

We see that when $c_3>0$, nothing special happens, this situation is qualitatively same as the case of topological black holes in GR~\cite{Birmingham:1998nr}:  when $k=1$,
there exists the Hawking-Page phase transition, the black hole with small horizon radius is thermodynamically unstable with negative heat capacity, while it is stable with positive
heat capacity for large horizon radius; when $k=0$ and $-1$, the Hawking-Page phase transition does not happen and the black holes are always thermodynamical stable with positive
capacity.

 When $c_3<0$, however, some interesting things appear.  In Fig.~\ref{T5}, we show the behavior of the temperature of the black holes with $k=1$ when $c_3 =-0.1$ and $c_3=-0.3$, respectively, while in Fig.~\ref{F5} the corresponding free energies are plotted.

\begin{figure}[h]
\centering
\includegraphics[scale=0.7]{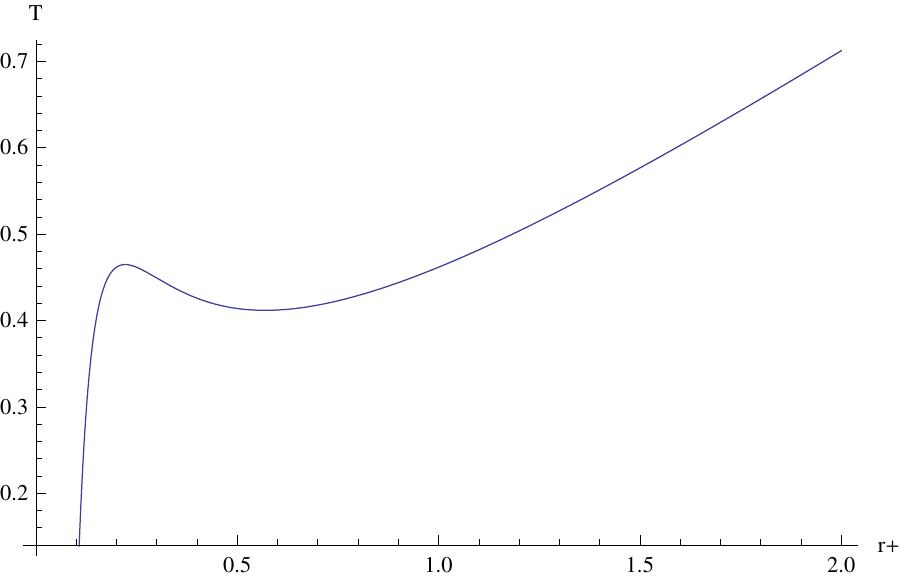}
\includegraphics[scale=0.7]{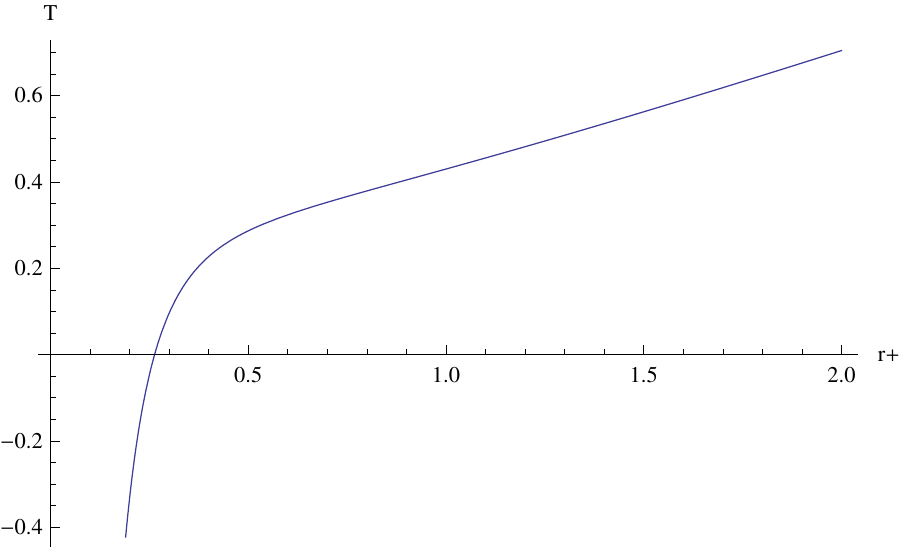}
\caption{The temperature $T$ of the black hole with $k =1 $ versus horizon radius $r_+$, here we take $l=1$.  Left: $c_3m^2=-0.1$ . Right: $c_3m^2=-0.3$  }
\label{T5}
\end{figure}

 From Fig.~\ref{T5} we see that the temperature behaviors are quite similar to the cases shown in Fig.~\ref{T(k=2)}, which implies that there exists a critical $c_{3crit}$. When $c_3 < c_{3crit}$, the black holes are always thermodynamically stable with a positive capacity, while $c_{3crit} <c <0$, there exist two stable branches with positive capacity, small horizon branch and large horizon
 branch, while for the intermediate  horizon, the black holes are thermodynamically unstable with negative capacity.  The heat capacity is given by
\begin{equation}
\label{capa5}
C_5=T\left(\frac{d S}{dT}\right )= \frac{6\pi^2 V_3 T_5 r_+^4}{\kappa^2 (-k +\frac{2r_+^2}{l^2}-\frac{2c_3m^2}{r_+})}.
\end{equation}
The critical $c_3$ is given by $c_{3crit} =-l/m^2\sqrt{54}$ for the case of $k=1$. From the free energy behavior shown in Fig.~\ref{F5},  we see that the first order phase transition
occurs when $c_{3crit} <c_3 <0$ and it disappears if $c_3< c_{3crit}$.

\begin{figure}[h]
\centering
\includegraphics[scale=0.5]{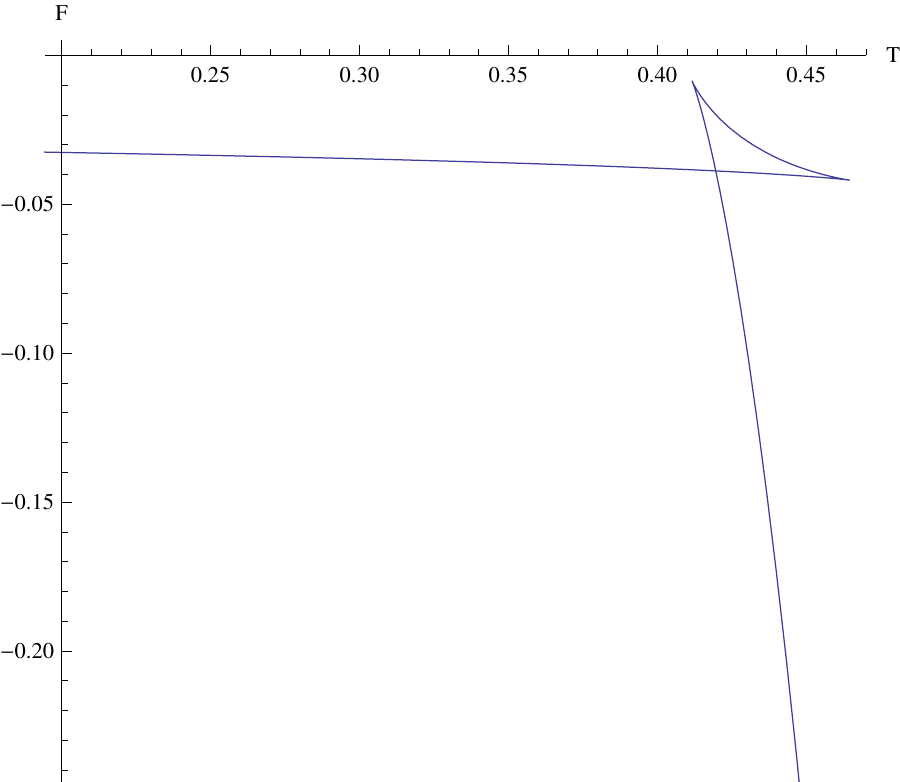}
\hspace{2.cm}
\includegraphics[scale=0.5]{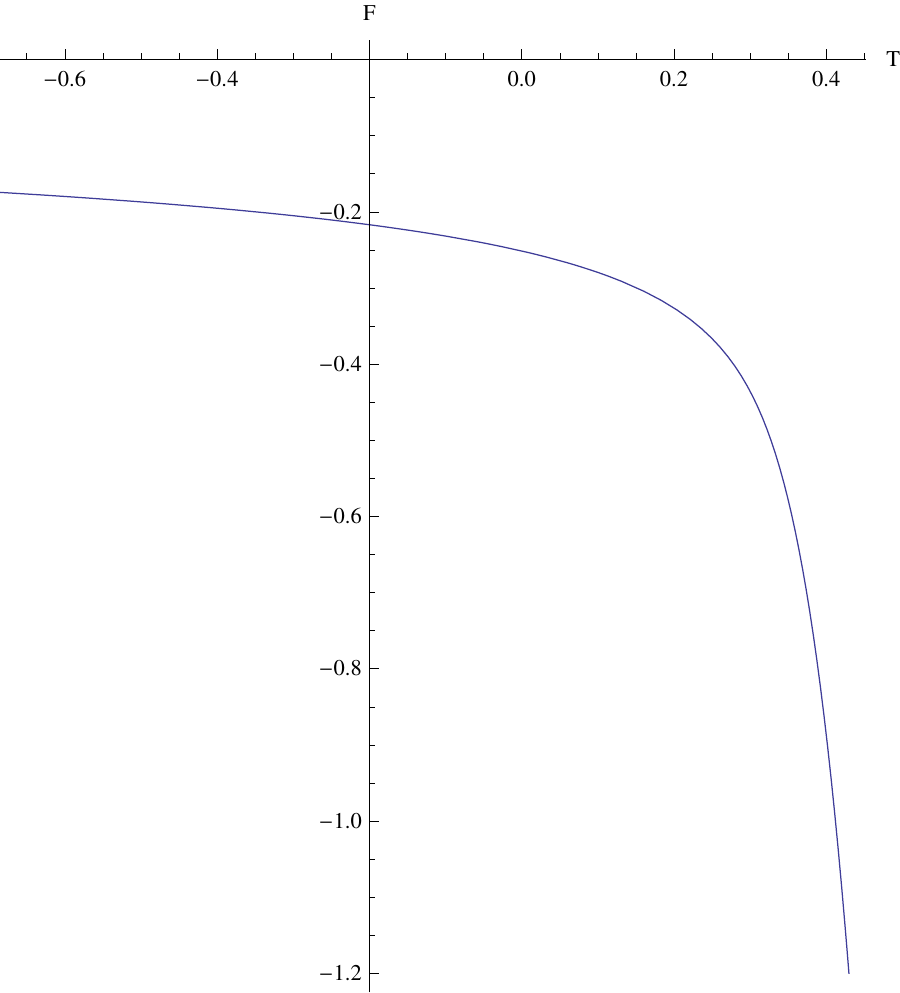}
\caption{The Helmholtz free energy of black hole  with $k =1 $ versus temperature $T$, here we take $l=1$.  Left: $c_3m^2=-0.1$. Right: $c_3m^2=-0.3$. }
\label{F5}
\end{figure}

Note that the first order phase transition discussed in the above occurs only in the case of $k=1$. In addition, it is easy to see from the temperature behavior (\ref{Temp}) that the term $c_4m^2$ in the six dimensional case can play the same role as the term $c_3m^2$ in the five dimensional case.

%%%%Conclusion%%%%%%%%

\section{Conclusions}
\label{sect:conclusion}

In this paper we have presented a class of charged black hole solutions in an $(n+2)$-dimensional massive gravity with a negative cosmological constant, and studied thermodynamics and
phase structure of the black hole solutions both in grand canonical ensemble and canonical ensemble in some details. The black hole horizon
can have a positive, zero or negative constant curvature.  In the massive gravity we have considered,
there are four terms in the potential associated with the graviton mass. In the four dimensional case, only two terms appear in the solution, while other two terms may occur in higher dimensional cases.
By using Hamiltonian approach, we have obtained conserved charges of the solutions and they satisfy the first law of black hole thermodynamics. It turns out the entropy of the black hole still obeys the area formula as in GR although the massive gravity is not diffemorphism invariant. In addition, we have shown that gravitational field equation at black hole horizon
can be cast into the first law form of black hole thermodynamics.

In the four dimensional case,  the black hole thermodynamics and phase structure crucially depend on the horizon curvature of the black holes and the sign of $c_2$.
In grand canonical ensemble,  the Hawking-Page phase transition happens for the case $k-\mu^2 /4 +c_2m^2 \ge 0$.  Namely it can appear only for the case $k=1$ if $c_2 <0$, while for the
 $k=0$ or $-1$ case,
the black holes are always thermodynamically stable with positive capacity.  When $c_2 >0$,  however, we have found that the Hawking-Page phase transition always can appear for any
horizon curvature if $k-\mu^2 /4 +c_2m^2 \ge 0$. In canonical ensemble, when the charge of the black hole is less than its critical one, a small/large black hole first order phase transition happens if $k +c_2m^2 \ge 0$. This phase transition behaves like the one in van de Waals system~\cite{Chamblin:1999tk} .

For the case in higher dimensional ($n+2 \ge 5$) case, we have found that even when the charge of the black hole is absent, the small/large black hole phase transition can appear, the coefficient $c_3m^2$ and (or) $c_4m^2$ can play the role as the charge does in the four dimensional case, if $c_3m^2$ and (or) $c_4m^2$ are (is) negative.  This is a remarkable result in massive gravity, which does not appear in GR.

Finally we mention here that the black hole solutions presented in this paper crucially depend on the choice of the reference metric (\ref{reference}). In general, if one can take the
 ansatz: $f_{\mu\nu}={\rm diag} (0,0, c_0^2 F(r)h_{ij})$, where $F(r)$ is a continuous function of $r$, we can also obtain the exact solution of the theory once $F(r)$ is specified. In this sense,
 the choice of the reference metric is an important issue in this class of massive gravity.

%\appendix

%\begin{acknowledgments}

\section*{Acknowledgments}
RGC thanks T. Kugo and N. Ohta for helpful discussions. This work was supported in part by the National Natural Science Foundation of China
(No.10821504, No.10975168, No.11035008 and No.11435006).
%\end{acknowledgments}

\end{document}